\documentclass[twocolumn, secnumarabic, amssymb, nobibnotes, aps, pre, showpacs, superscriptaddress]{revtex4}
\usepackage{graphicx}
\usepackage{epsfig}
\usepackage{float}
\usepackage{amsmath}
\begin{document}
\title{Fractional diffusion equation for aging and equilibrated random walks}
\author{V.~Yu.~Zaburdaev}
\email{vasily.zaburdaev@tu-berlin.de} 
\affiliation{{\it Institut f\"{u}r Theoretische Physik, Technische Universit\"{a}t Berlin, 
Hardenbergstr. 36, D-10623 Berlin, Germany}}
\author{I.~M.~Sokolov}
\email{igor.sokolov@physik.hu-berlin.de} \affiliation{{\it
Institut f\"{u}r Physik, Humboldt-Universit\"{a}t zu Berlin,
Newtonstr.15, D-12489 Berlin, Germany}}

\begin{abstract}
We consider continuous time random walks (CTRW) and discuss situations
pertinent to aging. These correspond to the case when the initial
state of the system is known not at preparation (at $t=0$) but at the 
later instant of time $t_1>0$ (intermediate-time initial condition). 
We derive the generalized aging diffusion equation
for this case and express it through a single memory kernel. The
results obtained are applied to the practically relevant case of the
equilibrated random walks. We moreover discuss some subtleties in the
setup of the aging subdiffusion problem and show that the behavior of
the system depends on what was taken as the intermediate-time initial
condition: whether it was coordinate of one particle given by
measurement or the whole probability distribution. The two setups lead
to different predictions for the evolution of a system. This fact
stresses the necessity of a precise definition of aging statistical
ensembles.
\end{abstract}
 \maketitle

\section{Introduction}
Fractional diffusion equations (FDE) nowadays can be considered as
an established and common mathematical tool with applications
widely distributed in natural and social sciences, biology, and
finance. Usually, one successfully uses them to explain anomalous
scaling behavior of some quantities of interest. However, when
finer details of anomalous transport processes come to a question,
such as aging for example, an investigator, equipped with FDE but
with the logic of classical diffusion, often fails to describe it
correctly. An intrinsically asymptotic character of FDE is more
crucial than in classical diffusion, since it involves nonlocal
time/space operators. Therefore, some particular tasks require the
knowledge of the dynamics underlying the FDE and not only their
final form.  In many cases, such dynamics can be modeled by
continuous time random walks (CTRW) \cite{MS84,MK00,BG90}. It
allows to follow all derivation steps in detail and obtain useful
analytical results. In the present work, we will also use CTRW as
a basis model.

The problem of aging is one of the fundamental questions relevant
in different fields of physics. The word ``aging'' essentially
denotes a whatever pronounced time-inhomogeneous behavior of a
physical system. It applies to all situations when the result of a
measurement depends explicitely on the time when this measurement
was performed (i.e. what is the time elapsed between the instant
when the system was prepared and the instant when the measurement
started). Aging and memory effects are intrinsic features of
glasses \cite{S78,B97,B06}, colloidal systems \cite{BKH03, ABM01},
granular materials \cite{OC03, JTMJ00}, diffusion processes in
random environment \cite{DMF99}, etc. Measurement setups and
physical reasons for such a time-inhomogeneous behavior can be
quite different, e.g. the physical system can age due to some slow
internal processes or due to external impacts. In what follows we
consider the CTRW model to demonstrate the nature and appearance
of aging effects which, in this case, can be completely
understood.

Different aspects of aging in continuous time random walks were
addressed e.g. in \cite{A1,A2,B03,BC03,SKB01}. The problem
discussed in the present article is related to the one of
Ref.~\cite{B03}, where the author considered the aging in the
subdiffusion process generated by a deterministic dynamical
system. We generalize the results of the Ref.~\cite{B03} and
explicitly express the generalized diffusion equation describing
aging CTRW by using a single memory kernel. Further, we consider
the situation of random walks with truncated power-law waiting
time distributions, which leads to equilibration.  We show that
the final form of the corresponding diffusion equation depends on
whether this waiting time distribution possesses the second moment
or not. In the latter case, the convergence of the probability
density function (PDF) for the concentration of particles to a
Gaussian is very slow.

We moreover stress that the results obtained in the
Ref.~\cite{B03} and generalized in the present work are pertinent
to a very \emph{specific} variant of the aging problem. To
demonstrate this, we discuss a seemingly similar (but in reality
very different) approach to aging. Although both approaches are
intimately related, they lead to different predictions for the
evolution of the system. We show that these approaches also
correspond to different experimental situations and discuss their
applicability and limitations.

The structure of the paper is as follows. In the Section
\ref{Barkai} we formulate the aging problem corresponding to the
setup of Ref.~\cite{B03}, introduce the generalized master
equation (GME) approach, and derive the corresponding fractional
diffusion equation for aging walks. We analyze the asymptotic form of the aging
diffusion equation and express it through a single memory kernel.
In the Section \ref{Equil}, we apply these results to the
equilibrated random walks. We derive the corresponding transport
equations with distributed order fractional time derivative and
discuss particularly interesting behavior of the moments of the
density distribution. In the Section \ref{GreenS}, we discuss another
approach to aging, the relevant experimental situation, and the
corresponding results. The last section (Section \ref{Conclusions})
is reserved for conclusions.

\section{Master equation for aging random walks}
\label{Barkai}

\subsection{General considerations}
Consider a system of particles created at $t_0=0$. After the system
was created, we let it evolve according to its internal dynamical laws
during an interval of time $t_1>0$. The initial distribution of
particles at preparation, $n(x,0)$ is unknown. At the time $t_1$ we
labeled some particles and thus created a known probability
distribution $n(x,t_1)$ of such marked particles (performed a
measurement). For example, we could irradiate our system with light
and create excited states of atoms or molecules of the interest inside
the region of the order of the light beam radius or its penetration
depth.  Equivalently, we may consider a single particle that has
started at time $t=0$ from an unknown location and at the time $t_1$,
was detected at the point $x_1$ and labeled.  The quantity of interest
is $n(x,t_2)$, the density at a given instant of time $t_2>t_1$, which
we assume to be experimentally accessible. Then the following question
can be put: If we only know the duration of the previous evolution,
$t_1$ (called \textit{aging time}), and the position of the particle
at $t=t_1$, what can we say about the future evolution of the particle's
position? How precise can we predict the position of this particle at
the time $t_2>t_1$, and how does this prediction depend on $t_1$ and
$t_2$?  In the case of many labeled particles, the same questions
apply to the profile at $t_2$ which can be obtained as a convolution
of the concentration profile at $t_1$ and the PDF of the
single-particle displacements. We note that the density of points in a
configuration space of some system can correspond not only to
coordinates of some real particles but to whatever other coordinates
characterizing e.g. temperature or magnetic field.  In our
explanation we, however, confine ourselves to the picture of particles.

\begin{figure}[h]
\leavevmode
\begin{center}
\epsfig{file=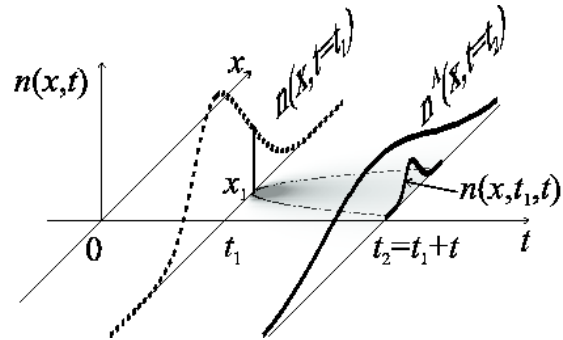,width=8.0cm} \caption{ The first aging
problem setup. The evolution starts at $t=0$ from an unknown
initial distribution. At $t=t_1$ the particles with a profile
$n(x,t=t_1)$ are labeled. At $t_2=t_1+t$ for the diffusion from a
single point we obtain the probability density $n(x,t_1,t)$ and
for the full profile the density $n^{A}(x,t=t_2)$. The dashed line
and the grayscale shadow in the $(x,t)$ plane are used to guide
the eye for the diffusion from a single point.\label{aging1}}
\end{center}
\end{figure}

The problem of aging in this setup can be considered as the
``intermediate-time initial condition'' problem: The system was
created at $t=0$ but the initial condition to the corresponding
transport equation is posed at a later instant of time $t=t_1$.  For
normal, Markovian diffusion this does not change the overall form of
the transport equation. In non-Markovian cases, especially in the ones
with long enough memory, it does, as we proceed to show. This aging
problem is schematically illustrated in the Fig.~\ref{aging1}. Assume
that the density of labeled particles at $t=t_1$ at point $x=x_1$ is
known. Starting from this point the particles diffuse, and at
$t_2=t_1+t$ acquire some distribution (not shown). By $n(x,t_1,t)$ we
denote our theoretical prediction for this distribution. In the case
when the whole profile $n(x, t=t_1)$ is known, the concentration
profile $n_{A}(x,t=t_2)$ at $t_2$ is given by a convolution of the
former one and $n(x,t_1,t)$.

For simplicity we start with the CTRW model on a one-dimensional
lattice; generalizations to the continuous case and to higher
dimensions are quite evident. Consider a discrete set of sites
marked by an index $i$.  By $n_{i}$ we denote the occupation
probability of each site (density of particles). After a certain
waiting time at the site $i$, a particle can jump to the two
neighboring sites $i-1, i+1$ with equal probability $1/2$. The
waiting time distribution at a site is governed by the probability
density $\psi(\tau)$. The properties of this function determine
the regime of diffusion. If the mean waiting time,
$\left<\tau\right>=\int\limits_{0}^{\infty}\tau \psi(\tau) d\tau$,
is finite, the resulting transport process will be a normal
diffusion. However, even in this case some aging effects can still
emerge, with the only exception of the exponential waiting time
distribution \cite{ZCh03}.  If the mean waiting time is infinite,
we are in the situation of anomalous subdiffusive behavior, where
the aging effects are the strongest.  It is known that in the
subdiffusive regime asymptotic transport equations have a form of
the fractional diffusion equation, where fractional
differentiation appears in the temporal part. It is easy to see
that in the case of the fractional time derivative, the solutions
of the corresponding equation do not posses the semi-group
property. This is a strong indication that the choice of the
starting point or an intermediate initial condition may noticeably
affect the following dynamics.

To consider this point in detail, we need to derive the transport
equation which adequately describes aging. Our approach is based
on the generalized master equation and is similar to the
phenomenological derivation of the diffusion or Fokker-Planck
equations using a combination of a continuity equation and the
equation for currents \cite{CGS}. The continuity assumption
contains essentially two balance conditions guaranteeing the
probability conservation: a local one (giving the balance between
the probability gain and loss at one site) and the one for
transitions between the two sites (representing particle
conservation during the jumps).  A balance equation at each site
reads
\begin{equation}
\dot{n}_{i}(t)=j_{i}^{+}(t)-j_{i}^{-}(t),
\label{sitebalance}
\end{equation}
where $j_{i}^{-}(t)$ is the loss current, i.e. the
probability for a particle to leave the site per unit time at time
$t$, and $j_{i}^{+}(t)$ is the gain current at a site.

A particle arriving to the site $i$ at time $t$ comes either from
the left or from the right. Probability conservation for
transitions between sites then reads
\begin{equation}
j_{i}^{+}(t)=\frac{1}{2}j_{i-1}^{-}(t)+\frac{1}{2}j_{i+1}^{-}(t).
\label{balance1}
\end{equation}
By combining (\ref{sitebalance}) and (\ref{balance1}) we get a
continuity equation
\begin{equation}
\dot{n}_{i}(t)=\frac{1}{2}j_{i-1}^{-}(t)+\frac{1}{2}j_{i+1}^{-}(t)-j_{i}^{-}(t)
\label{balance2}
\end{equation}
According to the waiting time distribution, the loss current at
time $t$ is connected to the gain current at the site at all
previous times: the particles which leave the site $i$ at the time
$t$ (making a step from $i$ to one of its neighbors) either were
at the same site $i$ from the very beginning, or came there at
some later time $0<\tau<t$. A probability density to make a step
at time $t$ when arriving at $\tau$ is given by the waiting time
distribution $\psi (t-\tau)$. Then for the loss current
$j_{i}^{-}(t)$ we can write:
\begin{equation}
j_{i}^{-}(t)=\psi
^{(1)}(t,t_{1})\widetilde{n}_{i,0}+\int_{0}^{t}\psi
(t-\tau)j_{i}^{+}(\tau)d\tau.\label{bal}
\end{equation}
However, for the aging initial conditions the probability to make
the first step after the measurement at $t_1$ is different from
$\psi(\tau)$. Here we postulated it to be independent of the
spatial position $i$ and denoted by $\psi^{(1)}(\tau,t_1)$. We
also assumed that the system was created at the time $t=0$, and a
measurement was performed at $t=t_1$. From here on we denote by
$t$ the time elapsed from $t_1$ and denote by
$\widetilde{n}_{i,0}\equiv\widetilde{n}_{i}(t_1)$ the initial
condition for the system's further evolution.
 By using
Eqs.(\ref{balance1}-\ref{balance2}) we can rewrite (\ref{bal}) as:
\begin{equation}
j_{i}^{-}(t)=\psi
^{(1)}(t,t_{1})\widetilde{n}_{i,0}+\int_{0}^{t}\psi
(t-\tau)\left[\dot{n}_{i}(\tau)+j_{i}^{-}(\tau)\right] d\tau.
\label{balance3}
\end{equation}
Now we have to find the expression for the forward waiting time
distribution of the first step $\psi^{(1)}(\tau,t_1)$ \cite{GL01}.
The forward waiting time $\tau$ is counted starting from the
observation point. Let us assume that the jump preceding $t_{1}$
(numbered $j-1$), took place at time $t_{j-1}=z$. The forward
waiting time distribution $\psi ^{(1)}(\tau,t_1)$ is
\begin{equation}
\psi ^{(1)}(\tau ,t_{1})=\int_{0}^{t_{1}}f(z)\psi (t_{1}-z+\tau
)dz, 
\nonumber
\end{equation}
where $f(z)dz$ is the probability to make a jump within the time interval
between $z$ and $z+dz$, so that $f(z)$ is the time-dependent density of steps.
This can be presented in the following form:
\begin{equation}
f(z)=\sum_{n=0}^{\infty
}f_{n}(z).
\label{summ}
\end{equation}
Here $f_{n}(t)$ is the probability density that it is exactly
$n$-th jump that takes place at time $t$. This one is given by an
$n$-fold convolution of the waiting time probability density $\psi
(t)$ with itself. Under Laplace transform with respect to $z$
Eq.(\ref{summ}) reads:
\begin{equation}
f_s=1+\psi_s+\psi^{2}_s+...=\frac{1}{1-\psi_s}.
\end{equation}
By indexes $p,s$ we will denote the Laplace components
corresponding to $t$ and $t_1$, and by $k$ the Fourier component,
respectively. The Laplace transform of $\psi^{(1)}(\tau,t_{1})$
with respect to $t_{1} $ is then:
\begin{equation}
\psi^{(1)}_s(\tau)=\frac{\text{e}^{s\tau }\left[
\psi_s-\int_{0}^{\tau }\text{e}^{-s\tau'}\psi(\tau')d\tau'\right]
}{1-\psi_s}. 
\nonumber
\end{equation}
The double Laplace transform of this
function with respect to both times $\tau$ and $t_1$ has
the following form:
\begin{equation}\label{phidouble}
\psi^{(1)}_{s,p}=\frac{\psi_s-\psi_p}{(p-s)(1-\psi_s)}.    
\end{equation}
With this information at hand, we take the Laplace transform of
(\ref{balance3}) with respect to $t$ to obtain:
\begin{equation}
j_{i,p}^{-}=\psi_{p}^{(1)}(t_{1})\widetilde{n}_{i,0}+\psi_{p}\left[
p\,n_{i,p}-\widetilde{n}_{i,0}+j_{i,p}^{-}\right].
\nonumber
\end{equation}
From the above equation we find the connection between loss
current and the occupation probability:
\begin{equation}
j_{i,p}^{-}=\frac{p\,\psi_p}{1-\psi_p}n_{i,p}+
\frac{\psi^{(1)}_{p}(t_{1})-\psi_p}{1-\psi_p}\widetilde{n}_{i,0}.
\label{Aged}
\end{equation}
Now we insert (\ref{Aged}) into the Laplace transform of
(\ref{balance2}) and convert it back to the time domain:
\begin{eqnarray}
\dot{n}_{i}(t) &=&\int_{0}^{t}\Phi (t-\tau)\left[
\frac{1}{2}n_{i-1}(\tau)+\frac{1}{2}n_{i+1}(\tau)-n_{i}(\tau)\right] d\tau  \nonumber \\
&+&\Phi^{(1)}(t,t_1)\left[
\frac{1}{2}\widetilde{n}_{i-1,0}+\frac{1}{2}\widetilde{n}_{i+1,0}-\widetilde{n}_{i,0}\right],
\label{GMEAged}
\end{eqnarray}
where $\Phi(t)$ and $\Phi^{(1)}(t,t_1)$ are defined through their
Laplace transforms:
\begin{equation}\label{phi}
\Phi_p=\frac{p\psi_p}{1-\psi_p},\quad\Phi^{(1)}_p(t_1)=
\frac{\psi^{(1)}_{p}(t_{1})-\psi_p}{1-\psi_p}.
\end{equation}

\subsection{The second memory term}
\label{Sec22}
We note now that Eq.(\ref{GMEAged})
can be rewritten in the form 
\begin{eqnarray}
\label{GMEAged2Kernel}
\frac{d}{dt} n(x,t) &=& \frac{d}{dt} \int_{0}^{t}
M(t-\tau) \times \\ 
&& \times \left[\frac{1}{2}n_{i-1}(\tau)
+\frac{1}{2}n_{i+1}(\tau)-n_{i}(\tau)\right]d\tau
\nonumber \\
&+&\Phi^{(1)}(t,t_1)\left[
\frac{1}{2}n_{i-1}(t_1)+\frac{1}{2}n_{i+1}(t_1)-n_{i}(t_1)\right],
\nonumber
\end{eqnarray}
where the memory kernel $M(t)$ is the inverse Laplace transform of 
the function $M_p = \psi_p/(1-\psi_p)$, which is connected with the density
of steps $f(t)$ (having the Laplace representation $f_p=1/(1-\psi_p)$) via
$M(t)=f(t)-\delta(t)$. Thus, $M(t)$ is the density of all steps excluding the first one.
Evidently, the asymptotic properties of the functions $M$ and $f$ are the same.

Now we concentrate on the temporal
asymptotic behavior of the source term. By using Eqs.
(\ref{phidouble}) and (\ref{phi}), for the double Laplace
transform of $\Phi^{(1)}(t,t_1)$ we can write:
\begin{eqnarray}
\label{Phi1ps}
  \Phi^{(1)}_{p,s} &=&\frac{1}{p-s}\left[\frac{1}{1-\psi_s}
-\frac{1}{1-\psi_p}\right]-\frac{\psi_p}{s(1-\psi_p)}
\\
   &=&\frac{1}{p-s}\left[\frac{1}{1-\psi_s}-\frac{1}{1-\psi_p}\right]
-\frac{1}{s(1-\psi_p)}+\frac{1}{s}.\nonumber
\end{eqnarray}
In order to understand the structure of the above expression we
use one remarkable property of the Laplace transform. For any
function $h$, such that its Laplace transform exists, the
following property for double Laplace transform holds
\cite{ZCh03}:
\begin{equation}
[h(t+t_1)]_{p,s}=\frac{h_p-h_s}{s-p},
\end{equation} 
where $s$ and $p$ are Laplace space variables corresponding
to $t$ and $t_1$ respectively. By using this fact we can rewrite
the expression for $\Phi^{(1)}(t,t_1)$ (\ref{Phi1ps}) in time domain:
\begin{equation}
    \Phi^{(1)}(t,t_1)=f(t+t_1)-f(t)+\delta(t),
\end{equation}
where $f(t)$ is the density of steps given by the inverse Laplace transform of
$1/(1-\psi_p)$. This one, as we have already seen, is connected with the first memory kernel
$M(t)$ via $M(t)=f(t)-\delta(t)$, so that it can be rewritten in the form
$\Phi^{(1)}(t,t_1)=M(t+t_1)+\delta(t+t_1)-M(t)$. However, for whatever aged system ($t_1>0$) the delta
function vanishes, so that the final result 
\begin{equation}
    \Phi^{(1)}(t,t_1)=M(t+t_1)-M(t)
\end{equation}
follows. This is a new expression which connects the source term with
the memory kernel of the generalized master equation for aging continuous time
random walks (\ref{GMEAged2Kernel}), which thus gets the form
\begin{eqnarray}
\label{GMEfinal}
&& \frac{d}{dt} n_i(t) = \\
&& \frac{d}{dt} \int_{0}^{t}
M(t-\tau) \left[\frac{1}{2}n_{i-1}(\tau)
+\frac{1}{2}n_{i+1}(\tau)-n_{i}(\tau)\right]d\tau
\nonumber \\
&&+\left[ M(t+t_1)-M(t) \right]\left[
\frac{1}{2}n_{i-1}(t_1)+\frac{1}{2}n_{i+1}(t_1)-n_{i}(t_1)\right],
\nonumber
\end{eqnarray}

In the more general continuous case, when the jumps' lengths are no more
discrete but rather distributed according to
some probability density $g(x)$, Eq.(\ref{GMEfinal}) turns to
an integral equation of the form:
\begin{eqnarray}
\label{GMEAgedIntegral2}
&& \frac{d}{dt} n(x,t) = \\
&& \frac{d}{dt} \int_{0}^{t} M(t-\tau)\int\limits_{-\infty}^{+\infty}g(y)\left[n(x-y,\tau)-n(x,\tau)\right]dy d\tau  \nonumber\\
&& + \left[ M(t+t_1)-M(t) \right] \int\limits_{-\infty}^{+\infty}g(y)\left[n(x-y,t_1)-n(x,t_1)\right]dy.
\nonumber
\end{eqnarray} 
The discrete case, Eq.(\ref{GMEfinal}) corresponds to the choice
$g(x)=[\delta(x-a)+\delta(x+a)]/2$ with $a$ being the lattice spacing.
The generalized master equations, Eq.(\ref{GMEfinal}) and
Eq.(\ref{GMEAgedIntegral2}), contain the standard random walk part
represented by the first term and an additional term representing the
memory on the initial conditions. This memory term is expressed
through the same CTRW memory kernel as encountered in the first term
on the right hand side and vanishes when either $t_1=0$, or
$\psi(\tau)$ is an exponential.

\subsection{Asymptotic form of the aging equation}
\label{Asform}
Now we would like to obtain the asymptotic form of
the above transport equation corresponding to large values of $x$
and $t$ (or respectively small $k$ and $p$). Assuming $n_i(t)$ to change 
slowly enough as the function of the site number $i$ one can 
change to the continuous description introducing the density
$n(x,t)$ with $x=ia$. The difference operators in  Eq.(\ref{GMEfinal})
can then be considered as a discrete approximation to a Laplacian, so 
that the corresponding master equations takes the form of generalized
diffusion equation with the additional memory term
\begin{eqnarray}
\frac{\partial}{\partial t} n(x,t) &=& \frac{a^2}{2} \frac{\partial }{\partial t} \int_{0}^{t}
M(t-\tau) \Delta n(x,\tau) d\tau \nonumber \\
&& +\left[ M(t+t_1)-M(t) \right] \frac{a^2}{2} \Delta n(x,0)
\label{GenTra}
\end{eqnarray}
The same form follows from the more general continuous form, Eq.(\ref{GMEAgedIntegral2}),
provided the jump length distribution $g(x)$ has a finite second moment ($\left<x^2\right><\infty$).
The corresponding proof follows either the standard Kramers-Moyal procedure or
can easily be obtained as the small-$k$ expansion in the Fourier space. 

For exponential waiting time distribution $\psi(t)=t_0^{-1} \exp(-t/t_0)$ the memory kernel
$M(t)=1/t_0$ and the equation takes form of the ordinary diffusion equation 
\begin{equation}
\frac{\partial}{\partial t} n(x,t) = \frac{a^2}{2 t_0}\Delta n(x,t)
\end{equation}
lacking the second memory term. The combination $K=a^2/2t_0$ is the usual diffusion
coefficient of the process. Another common choice of the waiting time
distribution (determined by both, real practical
situations and mathematical convenience) is a power law function,
e.g. $\psi(t)=\gamma/t_0(1+t/t_0)^{1+\gamma}$, with $0<\gamma<1$, where $t_0$ again
gives us the characteristic temporal scale of waiting times.
The expansion of its Laplace transform for a small
$p$ is: $\psi_p=1-\Gamma(1-\gamma)t_0^{\gamma}p^{\gamma}+O(p)$, and the corresponding
memory kernel $M(t)$ is given by 
\begin{equation}
M(t) \simeq \frac{1}{\Gamma(\gamma) t_0^\gamma} \frac{t^{\gamma-1}}{\Gamma(1-\gamma)},
\end{equation}
in which we recognize an expression proportional to the integral kernel of the
fractional derivative.
After substituting this expansion into (\ref{GenTra}) and some
algebra we obtain:
\begin{eqnarray}
\label{AgeEqFin}
&& \frac{\partial}{\partial t} n(x,t)= K_\gamma \;_0D^{1-\gamma}_{t}\triangle
n(x,t) \\ 
&& + K_\gamma \frac{1}{\Gamma(1-\gamma)}\left[\frac{1}{(t+t_1)^{1-\gamma}}
- \frac{1}{t^{1-\gamma}}\right] \triangle n(x,t_1). \nonumber
\end{eqnarray}
where the generalized diffusion coefficient $K_\gamma$ stands for
$K_\gamma = a^2/[2 \Gamma(\gamma) t_0^\gamma]$.  We must note that the
second memory term in Eq.(\ref{AgeEqFin}) contains the Laplacian,
whereas the corresponding term in the result of Ref.\cite{B03} does
not. The probable reason for that is an insufficient number of terms
retained in the asymptotic expansion in Ref.\cite{B03}.

\section{Equilibrated random walks}
\label{Equil} 
\begin{figure}[h]
\leavevmode
\begin{center}
\epsfig{file=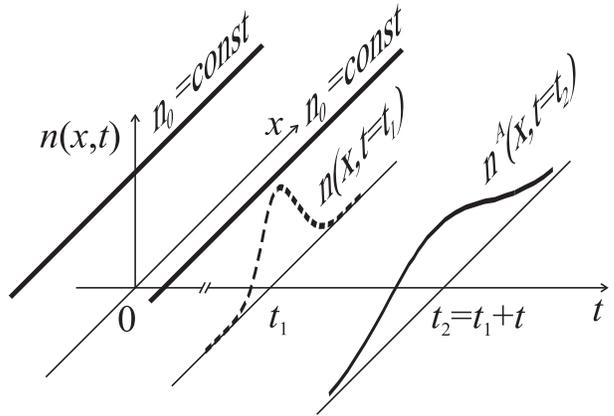,width=8.0cm} \caption{Problem setup for the
equilibrated random walks. At $t=t_1$ we
label some of the particles from their uniform distribution
and create the profile $n(x,t=t_1)$. Later we follow only the diffusion of those marked
particles and obtain at $t_2=t_1+t$ the density profile
$n^{A}(x,t=t_2)$. \label{Fig4}}
\end{center}
\end{figure}

Let us now consider the system which relaxes to a
true equilibrium. Such a system would correspond to CTRW with the
waiting time distribution possessing the first moment,
$\left<\tau\right> = \int_0^\infty \tau \psi(\tau)d\tau$, which is
however so large that the intermediate power-law asymptotics of
the function is still seen. The examples are a theta-truncated
power-law $\psi(\tau) = A/(1+\tau)^{1+\alpha} \theta(T-\tau)$ or
an exponentially truncated one $\psi(\tau) = A/(1+\tau)^{1+\alpha}
\exp(-\tau/T)$ (A is the normalization constant), or an
exponentially truncated one-sided Levy-distribution with the
Laplace transform
$\psi_p=\exp[-A(p+1/\lambda)^\alpha+A/\lambda^\alpha]$. Here $A$
is the parameter with the dimension $[\mathrm{T}]^\alpha$ and of
the absolute value of unity (the typical step time is set to one),
and $\lambda = (A \alpha/\tau)^{1/(1 - \alpha)}$, where
$\left<\tau \right>$ is the mean waiting time. The equilibrated
case is interesting because it is experimentally relevant: systems
are typically created much earlier than measurements are performed
and have enough time to equilibrate. The time lag between the
labeling particles at $t=t_1$ and final measurement at $t_2$ is
however small enough to probe nonequilibrium dynamics. The general
result for the aging problems can be easily applied to this
concrete example. The equilibrated case corresponds to the limit
$t_1 \rightarrow \infty$ of aging CTRW.  Analogously to the aging
problem, we illustrate the equilibrated problem setup in the Fig.
\ref{Fig4}.
The system prepared at $t=0$ evolves
during the time $t_1$, with $t_1\rightarrow \infty$ (in practice
it means that the system was created a long time before the
measurement took place and develops a uniform distribution of particles).
At the time $t=t_1$ we select some profile of the
particles $n(x,t=t_1)$ and follow only these particles until the
time $t_2=t_1+t$.

We start from the transport
equation (\ref{GenTra}) by taking its Fourier-Laplace
transform:
\begin{equation}\label{equil}
pn_{k,p}-\widetilde{n}_{k}(t_1)=-\frac{p\psi_p}{1-\psi_p}\frac{a^2}{2} k^2n_{k,p}
-\frac{a^2}{2} k^2\Phi^{(1)}_{p}(t_1),
\end{equation}
where we returned to our earlier notation for the second memory kernel $\Phi_1$.
In the above equation, the waiting time distribution has a finite
first moment, $\left<\tau\right>$, and therefore
$\psi_p=1-\left<\tau\right> p+o(p)$. In addition, the limit $t_1
\rightarrow \infty$ should be applied. In this case the expression
for the second memory kernel $\Phi^{1}(t)$ is simpler than before.
Its Laplace transform is given by (see (\ref{phidouble}) and
(\ref{phi})):
\begin{equation}
\Phi^{(1)}_p =
\left[\frac{1}{p\left<\tau\right>}-\frac{1}{1-\psi_p}+1\right]
\end{equation}
We note that in the time domain we now have $\Phi^{(1)}(t)=
1/\left<\tau\right> -f(t) + \delta(t)$ and that $f(t)$ still has a
physical meaning of the (time-dependent) density of steps. For a
pure power-law CTRW this function monotonously decays to zero with
time, while for equilibrated walks it decays to
$1/\left<\tau\right>$. The transition between the both regimes
takes place at at time $t_c \simeq \left<\tau\right>$. However,
this does not mean that the second memory function vanishes at
times which exceed $t_c$: the difference $
1/\left<\tau\right>-M(t)$ still can behave as a pure power-law in
some cases. It is important that the second memory function is
governed by the subleading term in the expansion of
$1/(1-\psi_p)$, whose behavior depends on whether the function
$\psi(\tau)$ possesses the second moment (we shall call this
situation ``a sharp cut-off'') or does not.

In the case of a sharp cut-off and for a small $p$,
$\psi_p=1-\left<\tau\right> p + M_2 p^2 + o(p^2)$, where $M_2$ is
the second moment of waiting times. Now we have:
\begin{eqnarray}
\frac{1}{1-\psi_p}&=&\frac{1}{\left<\tau\right> p + M_2 p^2 +
o(p^2)}\nonumber\\ &=& \frac{1}{\left<\tau\right> p} [1 +
(M_2/\left<\tau\right>) p +o(p)]. \nonumber\\ &=&
\frac{1}{\left<\tau\right> p} + \frac{M_2}{\left<\tau\right>^2} + o(1)
\end{eqnarray}
This means that for small $p$
\begin{equation}
\Phi^{(1)}_p=1 - \frac{M_2}{\left<\tau\right>^2} + o(1),
\end{equation}
so that the function $\Phi^{(1)}(t)$ is integrable, with the
integral being equal to $1-M_2/\left<\tau\right>^2$. This in turn
means that $\Phi_1(t)$ decays faster than $t^{-1}$ and its exact
decay form depends on how many moments does  $\psi(\tau)$ actually
possess. We note that the existence of the second moment is
necessary for such a behavior. The overall fractional diffusion
equation for such a process in the long-time limit tends to a
normal diffusion equation of the type
\begin{eqnarray}
\frac{\partial n(x,t)}{\partial t} =
K \triangle n(x,t) + C \delta(t) \triangle \widetilde{n}_{0},
\end{eqnarray}
where $K$ stands for the diffusion coefficient
$K=a^2/\left<\tau\right>$ and the constant $C$ is given by
$C=(1-M_2/\left<\tau\right>^2)a^2$, i.e. to the ordinary diffusion
equation with some correction to the initial condition. This
correction vanishes only for the Markovian random walk process
with exponential waiting time distribution for which $M_2 =
\left<\tau\right>^2$.

In the case when the second moment does not exist, the situation can
be vastly different. In order to present this type of behavior let us
consider a special example of the waiting time distribution for which
we can convert Eq.(\ref{equil}) in the Fourier-Laplace domain into a
distributed-order fractional diffusion equation with a source in space
and time. Let us take $\psi_p$ in a form
\begin{equation}
\psi_p=\exp \left[ - \frac{p \left<\tau\right>}{1+\Lambda
(p\left<\tau\right>)^{1-\alpha}}\right].
\end{equation}
This function is completely monotonic, which, according to the
Bernstein's theorem, means that it is a Laplace transform of a
probability density function. To prove this it is enough to note
that $\psi_p$ has a form $e^{-h_p}$ with the function $h_p>0$,
$\left.h_p\right|_{p=0}=0$ and possesses a monotonous derivative
(see \cite{SCK04}). For $p \rightarrow 0$, $\psi_p$ behaves as
$\psi_p \simeq 1 - p\left<\tau\right>$, i.e. the corresponding PDF
$\psi(\tau)$ has the mean value of $\left<\tau\right>$. The
parameter $\Lambda$ is a free parameter of the order of unity
which governs the precise form of the crossover.

Provided that $p$ is small enough,  we can take $\psi_p \approx 1-
p \left<\tau\right>/[1+\Lambda (p\left<\tau\right>)^{1-\alpha}]$
which corresponds to $1/(1-\psi_p)= 1/ (p \left<\tau\right>) +
\Lambda (p\left<\tau\right>)^{-\alpha}$. By substituting this
expression into (\ref{equil}) we obtain:
\begin{eqnarray}
pn_{k,p}-\widetilde{n}_{0,k} &=& -K\left[1+\Lambda
(p\left<\tau\right>)^{1-\alpha}\right]
k^2 n_{k,p}\nonumber\\
&+& K\left<\tau\right>\Lambda(p\left<\tau\right>)^{-\alpha} k^2
\widetilde{n}_{0,k}.
\end{eqnarray}
In normal space-time domain this equation corresponds to
\begin{eqnarray}
\frac{\partial n(x,t)}{\partial t} &=&
K\left[1+\Lambda\left<\tau\right>^{1-\alpha}\;_0D_t^{1-\alpha}\right]
\triangle n(x,t)\nonumber\\
 &-& K\frac{\Lambda\left<\tau\right>^{1-\alpha}}{\Gamma(\alpha)}t^{-1+\alpha}
\triangle\widetilde{n}_{0},
\end{eqnarray}
Note the distributed order derivative ($1+\Lambda
\tau^{1-\alpha}\;_0 D_t^{1-\alpha}$) on the r.h.s. \cite{SCK04},
which is typical for systems showing crossover behavior, and the
extremely slowly dissolving of the initial condition!

The difference between the cases of the equilibrated random walks
corresponding to the waiting time distributions with or without
second moment (mirrored in the difference of the form of the
corresponding generalized Fokker-Planck equation) is intimately
connected with the question whether the mean forward waiting time
(the mean waiting time of the first step after the beginning of
observations) exists. Indeed, looking at the limit of the Laplace
transform in the first argument of the forward waiting time
distribution for the equilibrated case ($t_1 \to \infty$) we see
that $\psi_p^{(1)} \rightarrow (1-\psi_p)/\left<\tau\right> p$.
Therefore, the first moment of $\psi^{(1)}(t)$, $\left\langle t_1
\right\rangle = \int_0^\infty \psi^{(1)}(t,\infty) t dt =
-\frac{d}{dp} \left. \psi_p^{(1)}\right|_{p=0}$ is equal to $M_2/2
\left<\tau\right>^2$ provided $M_2$, the second moment of the
waiting time distribution, exists, and diverges otherwise.

One of the interesting peculiarities of the equilibrated random
walks is found in the behavior of its moments. It is known that
the second moment of equilibrated CTRW behaves diffusively at all
times. However, the higher moments of the density profile are much
more exotic when the second moment of $\psi(\tau)$ does not exist.
From Eq.(\ref{equil}) we obtain:
\begin{eqnarray}
n_{k,p} &=& \frac{1}{p} - \frac{1-\psi_p}{p^2 \left<\tau\right>}
\frac{a^2 k^2}{1-\psi_p+ \psi_p a^2 k^2/2} \nonumber \\
&\approx&  \frac{1}{p} - \frac{1-\psi_p}{p^2 \left<\tau\right>}
\frac{a^2 k^2/2}{1-\psi_p + a^2 k^2}\nonumber \\
&=& \frac{1}{p} \left[ 1 - \frac{1}{p\left<\tau\right>} 
\frac{a^2 k^2}{2} + ...  \right. \nonumber \\ 
&& \left. + (-1)^{2n}\frac{1}{p\left< \tau \right>} \frac{a^{2n} k^{2n}}
{(1-\psi_p)^{n-1}}+...\right]. 
\end{eqnarray}

Now we easily find the even moments of the
density by the following formula:
$M_{2n}=\left.(-1)^n(d^{2n}/dk^{2n})n_{k}\right|_{k=0}$. We thus
have:
\begin{eqnarray}
& M_{0,p}=1/p & M_0(t)=1  \nonumber \\
& M_{2,p}=2K p^{-2} & M_2(t)=2K t  \nonumber \\
& M_{4,p}=\frac{24 K^2 \left<\tau\right>}{p^2 (1-\psi_{p})} & M_4(t)= 24 K^2 \left<\tau\right> f_4(t)  \nonumber \\
&...& ... \\
& M_{2n,p}=\frac{(2n)!K^n \left<\tau\right>^{n-1}}{p^2
(1-\psi_p)^{n-1}} & M_{2n}(t)= (2n)! K^n \left<\tau\right>^{n-1} f_{2n}(t). \nonumber
\end{eqnarray}

All odd moments are zero due to the the symmetry of the case
considered here. The functions $f_{2n}(t)$ are the inverse Laplace
transforms of $f_{2n,p}=1/p^2(1-\psi_p)^{n-1}$. For the waiting
time distribution with infinite second moment considered above,
its asymptotic Laplace transform is as follows:
\begin{equation}
f_{2n,p}=\frac{1+\Lambda(p\left<\tau\right>)^{1-\alpha}}{p^{n+1}\left<\tau\right>^{n-1}}
\end{equation}
In the time domain it corresponds to the following behavior:
\begin{equation}
f_{2n}(t) \simeq \left\lbrace
\begin{array}{lll}
\displaystyle
\frac{\Lambda^{n-1}\left<\tau\right>^{-\alpha(n-1)}}{\Gamma[(n-1)\alpha
+2]}
t^{(n-1)\alpha +1}& \mathrm{for} & t < \left<\tau\right> \\
\displaystyle
\frac{1}{n!\tau^{n-1}}t^n & \mathrm{for} & t \gg \left<\tau\right>
\end{array}
\right.
\end{equation}
This means, for example, that the distribution of the particles'
positions at an intermediate time is considerably platykurtotic,
and only slowly tends to a Gaussian in course of time with a
typical transition time $\left<\tau\right>$. We note here that such a slow
convergence to a Gaussian is typical also for other random walks
models with truncated power-laws, as exemplified by truncated
L\'evy flights \cite{MaSt,SCK204}.

\section{Green's function approach and the problem of aging ensembles}
\label{GreenS}

In the previous sections, we considered the results for a specific
setup of an aging problem corresponding especially to a
displacement of a single particle labeled at $t=t_1$ in course of
its further temporal evolution or of the ensemble of such labeled
particles. Let us consider another setup for aging problem. Assume
that the \textit{full profile} of the density of particles (or the
full probability density for a single particle) at $t=t_1$ is
known (for example, we started with a thermodynamically large
ensemble of particles and were able to measure their density at
$t=t_1$ with high enough precision) or is known \textit{in
principle} as it is assumed when calculating, say, the correlation
functions \cite{BarMuk}.  Does it give us additional information
which might improve the prediction of the particles' positions?
What would be the following evolution of this profile and how will
it depend on the aging time?

The GME approach used in Sec. \ref{Barkai} relies on a special
definition of the aging ensemble. The system was created at $t=0$
and evolves according to the CTRW dynamics up to the time $t_1$,
where the first measurement of the particle's position (or the the
particles' positions) takes place. No explicite notion about the
initial distribution of the particles' positions at $t=0$ is
assumed or used. In any case, we follow up only the particles
tagged at $t=t_1$. The situation considered in the present section
corresponds to a {\em different ensemble}.

Imagine now that we have two experiments (probably with different
initial conditions), which are such that the exact measured
density in one of them is the same as the density of labeled
particles in the another one. Will the further development of the
densities be the same? If not, what is the difference? As we
proceed to show, the two situations lead to {\em different
results} for the higher moments of the distribution. It is worth
mentioning that the distribution of the first exit times after the
measuring event is the only quantity affected by aging and it
determines the future evolution of the system for both definitions
\cite{B03}. It depends on the initial conditions through the
correlation between the particle's position and the forward
waiting time, as discussed in Ref. \cite{BS}.

\begin{figure}[h]
\leavevmode
\begin{center}
\epsfig{file=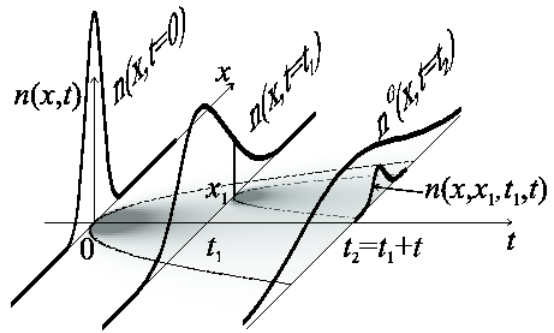,width=8.0cm} \caption{ The second aging
problem setup. At $t=t_1$ the full density profile, $n(x,t=t_1)$
is measured. By using the Green's function it is possible to
propagate it back and reproduce the initial distribution at $t=0$,
$n(x,t=0)$. Then it is forwarded until the time $t_2=t_1+t$ to
obtain the prediction for the future evolution $n^{G}(x,t=t_2)$.
In contrast to the first aging problem, diffusion from a single
point $x_1$ will depend on both, the starting point $x_1$ and the
aging time $t_1$: $n(x,x_1,t_1,t)$. The grayscale shadows and
dashed line in the $(x,t)$ plane show the diffusion from the
initial distribution and a single point.\label{aging2}}
\end{center}
\end{figure}

The difference between the two setups, the one of
Fig.~\ref{aging1} and the one of Fig.~\ref{aging2}, on the
qualitative level, can be understood already now. In the given
point $x_1$ at the time $t=t_1$, there are particles which have
arrived different times ago and, therefore, will make their next
steps also at different times \cite{Ch95,ZCh02}. In the first
aging setup, we virtually ignore the dependence of these exit
times on $x_1$ (the relative position in the full density profile,
if the latter is known). Therefore, we assume that in all points
the distribution of exit times depends only on $t_1$. In the
second aging problem, we take into consideration the relative
position of the point $x_1$. It is clear that particles which
occur to be found at the wing of the density profile typically
have made much more steps than those found close to the center of
the profile. Moreover, the particles which had to make a lot of
jumps probably have arrived at their actual position very
recently, i.e. start their last waiting period not long ago before
$t_1$. This implies that the distribution of exit times depends on
$x_1$ which affects the diffusion of particles out of this point.
Therefore corresponding profiles in the Fig. \ref{aging1} and
\ref{aging2}, $n(x,t_1,t)$ and $n(x,x_1,t_1,t)$ are different. By
repeating the same argumentation for each point in the profile
$n(x,t=t_1)$ we may conclude that $n^{A}$ and $n^{G}$ will be
different as well. We will come back to this question later in the
text.

Let us now return to the case when the PDF of particles' positions
at $t=t_1$ is known exactly and note that the temporal evolution
of the PDF from true initial conditions to the state at time $t$
is described by the linear evolution operator $\hat{S}(t)$.
If the time evolution operator has an
inverse defined at least on a set of relevant PDFs $n(x,t)$, one
can propagate this PDF back in time to $t=0$, thus getting
$n(x,0)=n_{0}(x)$ and then forward in time up to $t_2>t_1$ to
obtain $n(x,t_2)$ \cite{Barsegov}:
\begin{equation}
n(x,t_2)=\hat{S}(t_2)\hat{S}^{-1}(t_1) n(x,t_1).
\end{equation}
This defines the linear time evolution operator
$\hat{S}(t_2;t_1)=\hat{S}(t_2)\hat{S}^{-1}(t_1)$ which gives us
the PDF at the time $t_2$ provided the PDF at the time $t_1$ is
known.

Let us first find the explicit form of the $\hat{S}(t_2;t_1)$
operator and then discuss the difference between this aging
problem, and the first one, considered above. The time evolution
operator, $\hat{S}(t_2;t_1)$,  follows from the explicit form of
the solution of the transport equation via the Green's functions
method: it is the integral operator containing the Green's
functions of the standard CTRW equation, $G(x,x',t)$.
Let us recall the form of the transport
equation for CTRW without aging, i.e. all particles were
introduced at $t=0$ without history (the first part of
Eq.(\ref{GMEAged}) or Eq.(\ref{GMEAgedIntegral2})):
\begin{equation}
\dot{n}(x,t)=\int_{0}^{t}\Phi
(t-\tau)\int\limits_{-\infty}^{+\infty}g(y)\left[
n(x-y,\tau)-n(x,\tau)\right]dy d\tau.
\label{standard}
\end{equation}
The Green's function is defined as the solution of the above
equation with $n(x,t=0)=\delta(x-x')$. Then, for arbitrary initial
distribution $n_{0}(x)$ we can write:
\begin{equation}
n(x,t)=\hat{S}(t) n(x,0) = \int_{-\infty}^{+\infty}
G(x,x',t)n_{0}(x')dx'.
\end{equation}
For the homogeneous situation considered here, the Green's
function depends only on the difference of its spatial variables:
$G(x,x',t)=G(x-x',t)$. Therefore, the solution can be found as a
convolution of the initial condition with the Green's function. By
denoting the convolution operation with *, we can write
$\hat{S}(t)= G(x,t)*$, i.e.
\begin{equation}
n(x,t)=\hat{S}(t)n_{0}(x)\equiv G(x,t)*n_{0}(x).
\end{equation}
The PDFs of the particles' distributions at time $t_1$ and
$t_2=t_1 + t$  are thus given by
\begin{eqnarray}
n(x,t_1+t)&=&G(x,t_1+t)*n_0(x),\nonumber\\
n(x,t_1)&=&G(x,t_1)*n_0(x)
\end{eqnarray}
Under the Fourier transforms, convolutions are changed into simple
products of Fourier components:
\begin{eqnarray}
n_k(t_1+t)&=&G_k(t_1+t_2)n_{0,k},\nonumber\\
n_k(t_1)&=&G_k(t_1)n_{0,k}.
\end{eqnarray}
From these two we formally find:
\begin{equation}
n_k(t_1+t)= \frac{G_k(t_1+t)}{G_k(t_1)}\cdot n_k(t_1),
\label{Green_answ}
\end{equation}
and now write the time evolution operator $\hat{S}(t_2,t_1)
=\hat{S}(t_2)\hat{S}^{-1}(t_1)$ as an integral convolution
operator $T(x,t_2,t_1)*$ with the integral kernel being the
inverse Fourier transform of $G_k(t_1+t)/G_k(t_1)$.

For the CTRW model its Green's function can be easily found
analytically \cite{MK00}, so that also the corresponding operator
$\hat{S}(t_2,t_1)$ can be immediately calculated. The Green's
function for a CTRW equation (the solution of
(\ref{standard})) in the Fourier-Laplace space is
\begin{equation}
G_{k,p} = \frac{1-\psi_p}{p\left(1-\psi_pg_{k}\right)}. 
\end{equation}
We can now compare the resulting density distributions obtained in
the two aging problems, namely Eq.(\ref{GMEAgedIntegral2}) and
(\ref{Green_answ}). To do this we analyze the difference of the
two densities $\delta n(x,t)=n^{G}-n^{A}$, with $n^{G}$ being the
PDF predicted by the Green's function approach (\ref{Green_answ}),
and $n^{A}$ being the one given by the by the probabilistic
approach (\ref{GMEAgedIntegral2}) of Section \ref{Sec22}. In particular we
concentrate on the behavior of moments of the density difference
defined as $\left<x^n\right>_{\delta
n}=\int_{-\infty}^{\infty}x^{n}\delta ndx$.

As an example we consider the following situation. Imagine that
the system evolved starting from a sharp (delta-function) initial
condition $n(x,t=0)=\delta(x)$. In this case the particles'
distribution at $t=t_1$ is given by the inverse Fourier transform
of $n_k(t_1)=G_k(t_1)$. This distribution is now used as the
intermediate-time initial condition at $t=t_1$ in both approaches,
in the one of Sec. \ref{Barkai} and in the one of the present
Section.  Then the results of these two approaches are compared.
Eq.(\ref{Green_answ}) reduces in this case to a simple formula
$n^{G}_k(t+t_1)=G_k(t+t_1)$.

Omitting the details of rather tedious calculations, we can show that
the second moment of the density difference, $\left<x^2\right>_{\delta
n}$, is equal to zero.  However, already the next even moment,
$\left<x^4\right>_{\delta n}$, deviates from zero. The final
expression for the forth moment of the density difference reads:
\begin{eqnarray}
\label{x4}
\left<x^4\right>_{\delta
n}&=24&\frac{\sin\pi\gamma}{\pi\gamma\Gamma(\gamma)^2}
\int\limits_{0}^{t}\frac{d\tau}{(t-\tau)^{1-\gamma}} \\
&\times&\int\limits_{0}^{t_1}\left(\frac{\tau_1}{\tau}\right)
^{1-\gamma}\frac{1}{\tau_1+\tau}
\left[t_{1}^{\gamma}-(t_1-\tau_1)^{\gamma}\right]d\tau_1.\nonumber
\end{eqnarray}
In Fig.1 we plot $\left<x^4\right>_{\delta n}$ given by (\ref{x4}) as a function of
time for three different aging times, for the case $\gamma=1/2$
(the exponent in the power tail of the waiting time distribution).
\begin{figure}[h]
\leavevmode
\begin{center}
\epsfig{file=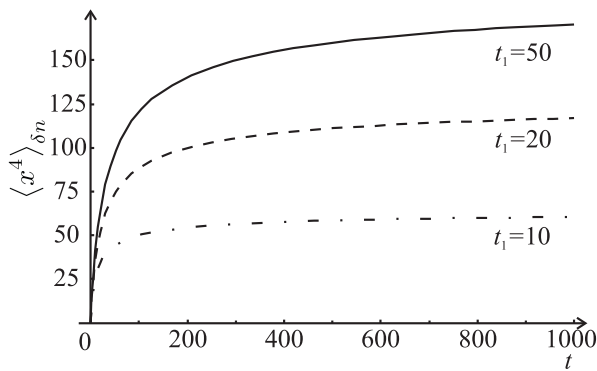,width=8.0cm} \caption{The forth moment of
$\delta n(x,t)=n^{G}-n^{A}$, as a function of time, $t$, for
different aging times $t_1$ and $\gamma=1/2$. Its deviation from
zero shows the difference between the prediction given by Eq.(\ref{GMEAgedIntegral2}) 
as compared to the Green's function result
(\ref{Green_answ}). \label{moment}}
\end{center}
\end{figure}
These findings show that the approach based on the
backward-forward propagation delivers a result which is different
from the one given by the approach of Sec. \ref{Barkai}. It is
also necessary to stress, that the aged propagator $n(x,t,t_1)$
given by (\ref{GMEAgedIntegral2}) \emph{is not} the Green's function of
aged anomalous diffusion, since it does not reproduce the PDF of
the particles exactly. The Gaussian situation showing no aging is
the only one when the both approaches are equivalent (and give the
same result): in this case $G_k(t)\propto\exp(-Dk^2t)$ so that
$G_k(t_1+t)/G_k(t_1)=G_k(t)$.

Let us discuss some implications of the two approaches. The
Green's function approach utilizes more information about the
intermediate stage of the system (the full knowledge of the whole
distribution) and corresponds to a different nonequilibrium
statistical ensemble than the one of Sec.~\ref{Barkai}. On the
other hand, the approach of Sec.~\ref{Barkai} assumes no knowledge
about the initial condition, i.e. essentially starts from the
assumption that at $t=0$ the distribution of particles in space
was uniform (which is the most reasonable assumption about the
state of the statistical system provided no additional information
is given and corresponds exactly to the Jaynes' information
approach to statistical mechanics \cite{Jaynes}). It is clear that
the homogeneous distribution does not evolve in course of the time
and stays uniform up to the time $t=t_1$. At $t=t_1$ we choose
some profile $n(x,t=t_1)$ from this distribution (see the Fig.
\ref{Fig4}) and follow its evolution.

As we have mentioned above, the physical reason for being able to
predict exactly the dynamics of particles after a measurement, is
the precise knowledge of their exit times. The Green's function
approach uses this information only indirectly, just propagating
the profile back and then forward and automatically provides the
corresponding distributions.  This approach is not literally
applicable in a whatever case when the intermediate time density
distribution is not a full profile resulting from the previous
evolution of some initial density. There exist, however methods
which work with microscopic distributions directly and can be
applied to such cases as well. The first one is based on the
generalized transport equation with microscopic details taken into
account \cite{ZCh03}, another one considers two-point probability
distributions \cite{BS, BauleEPL}. Both of these more general
approaches (not restricted to the aging problems) give answers
corresponding to the Green's function approach when starting from
a concentrated initial condition (and therefore for a whatever
exactly known one). Moreover, these approaches allow for the full
solution of the aging problem in the setup of the
Sec.~\ref{Barkai}, where the exit times after the measurement tend
to be independent of coordinate. The results for the equilibrated
walks provided by methods of Refs. \cite{ZCh03,BS} and by that of
the Sec.~\ref{Barkai} coincide since the information about the
initial state of the system is forgotten.

It is important to state that the two ensembles corresponding to
the two approaches discussed in this work are not the only two
alternatives, but two of a quite broad spectrum of possibilities.
They provide however the two limiting situations: the full
knowledge and no knowledge about this distribution. There are
various situations possible, when the distribution of
the particles' positions at the intermediate time is known to some
extent or with some uncertainty, in which case the results for
aging will depend on what exactly was measured at $t=t_1$ and what
the precision of this measurement was.

\section{Conclusions}
\label{Conclusions} 
Summarizing our findings, we have derived a general equation for the
PDF of the particle's position in aging CTRW, corresponding to the
``intermediate-time initial condition'' setup, in which the coordinate
of the particle is measured at some time $t_1$ after the preparation
of the system considered to take place at $t=0$.  We were able to
express the second memory function, describing the dissolving of the
initial condition, through the memory kernel of the CTRW equation. We discussed the ensuing forms of the transport equation for the case of CTRW with power-law
waiting time distributions lacking the first
moment, as well for the equilibrating situations, where the first
moment exists. We have shown that the exact form of the corresponding
equation for equilibrating walks depends on whether the second moment
of the waiting time distribution exists or not. In the first case, the
normal diffusion equation appears, in the second case, the evolution is
described by the diffusion equation with distributed-order temporal
derivative and with a very slowly dissolving second memory function following
the power law.  The asymptotic behavior of the higher moments of the
density profile for this case was also calculated and indicated a
slow convergence to the Gaussian behavior.

Moreover, as we tried to show, the problem of aging is a very delicate
task even in the framework of exactly solvable model of CTRW. To
illustrate this we have considered and solved another possible aging
setup corresponding to the full knowledge of the PDF of the particles'
positions at time $t_1$ after the preparation of the system at
$t=0$. The two setups give different predictions for the evolution of
the particles density and correspond to different experimental
realizations. This shows that there is still a room
for further developments in the area. One of those is the
question of the prediction of the evolution based on an incomplete
intermediate-time initial condition. This work is currently in
progress.

\begin{acknowledgments} We would like to thank
Profs. J.~Klafter and E.~Barkai for useful discussions. IMS
thankfully acknowledges the financial support by DFG within the
SFB555 research project.

\end{acknowledgments}



\begin{thebibliography}{99}
\bibitem{Barsegov} This is essentially the idea used in
Ref. \cite{BarMuk}.  However, the technical implementation of this
idea is faulty.  This can be seen when considering Eq.(18) of Ref.
\cite{BarMuk}.  We note that taking $t_2=t_1=t$ in this equation
has to give $W(X_2,t;X_1,t)=\delta(x_2-x_1)$: the particle cannot
be at two different points at the same time.  On the other hand we
note that due to the completeness of the system of eigenfunctions
of the corresponding Hermitian operators we have $\sum_n
\psi_n(X_1) \psi_n(X_2) = \delta(x_2-x_1)$. Due to the uniqueness
of the eigenfunction expansion this expression and Eq.(18) with
$t_2=t_1=t$ are compatible only in the case when
$E_\alpha(z)E_\alpha(-z)=1$, which can be proven wrong for
whatever $\alpha \neq 1$.  A trivial example is given by
$E_{1/2}(z)=e^{z^2}[1+\mbox{erf}(z)]$.  Thus the assumption that
backward propagator is the solution of the fractional backwards
Kolmogorov equation seems to be not a quite transparent choice in
all cases except for the trivial Markovian one where
$E_1(z)=\exp(z)$.
\end{thebibliography}
\end{document}